\documentclass{article}
\usepackage{arxiv}
\usepackage{color}
\usepackage{subfigure}
\usepackage{graphicx}
\usepackage{dcolumn}
\usepackage{bm}
\usepackage{floatrow}
\usepackage{amsmath}
\usepackage{makeidx}
\usepackage{epsfig}
\usepackage{array}
\usepackage{epstopdf}
\usepackage{multirow}
\usepackage[utf8]{inputenc} 
\usepackage[T1]{fontenc}    
\usepackage{hyperref}       
\usepackage{url}            
\usepackage{booktabs}       
\usepackage{amsfonts}       
\usepackage{nicefrac}       
\usepackage{microtype}      
\usepackage{lipsum}		
\title{Thermodynamic equilibrium of biological macromolecules under mechanical constraints}
\author{
  Ashkan Shekaari \\
  Department of Physics\\
  K. N. Toosi University of Technology\\
  Tehran, Iran \\
  \texttt{shekaari@email.kntu.ac.ir} \\
   \And
 Mahmoud Jafari \thanks{Corresponding author} \\
  Department of Physics\\
  K. N. Toosi University of Technology\\
  Tehran, Iran \\
  \texttt{jafari@kntu.ac.ir} \\}
\begin{document}
\maketitle
\begin{abstract}
Equilibrating proteins and other biomacromolecules is cardinal for molecular dynamics simulation of such biological systems in which they perform free dynamics without any externally-applied mechanical constraint, until thermodynamic equilibrium with the surrounding is attained. However, in some important cases, we have to equilibrate the system of interest in the constant presence of certain constraints, being referred to as {\em{constrained equilibration}} in the present work. A clear illustration of this type is a single amyloid $\beta$-strand or RNA, when the reaction coordinate is defined as the distance between the two ends of the strand and we are interested in carrying out replica-exchange umbrella sampling to map the associated free energy profile as the dependent quantity of interest. In such cases, each sample has to be equilibrated with the two ends fixed. Here, we introduced a simulation trick to perform this so-called constrained equilibration using steered molecular dynamics. We then applied this method to equilibrate a single, stretched $\beta$-strand of an amyloid beta dodecamer fibril with fixed ends. Examining the associated curves of the total energy and the force exerted on the practically-fixed SMD atom over the total timespan broadly supported the validity of this kind of equilibration.
\end{abstract}
\keywords{Constrained equilibration\and Biomacromolecules\and Replica-exchange umbrella sampling}
\section{Introduction}
Examining the underlying free energy profile (FEP) is still of foremost importance in fully understanding a large number of chemical processes such as protein-ligand binding~\cite{1}, or drug transportation across the cell membrane~\cite{2}, which cannot be accurately predicted, for instance in the field of computer-aided rational drug design~\cite{3}, without having the associated free energy behaviors~\cite{4}. A rich set of numerical tools for reliable determination of free energy changes based on the fundamental principles of statistical mechanics~\cite{5} has so far been developed and is now within reach. This so-called methodological set, in conjunction with recent advancements in computational power, has unprecedentedly enhanced the fields of free energy calculations, and therefore computational structural biology~\cite{6}, widening their areas of application as well.

The theoretical framework for free energy calculations was established a long time ago and several approximations have so far been accordingly developed~\cite{7,8,9,10,11,12}. However, central to accurate determination of free energy difference between initial and final states of a system is to sufficiently explore the configurational space of the initial state so that relevant, low-energy configurations of the final state could adequately be sampled. Conventional methods, particularly molecular dynamics (MD)~\cite{13} and Monte Carlo (MC)~\cite{14}, are not indeed successful in this respect~\cite{15} based on the fact that the system of interest could be trapped in a few conformational states (i.e., local minima) during the simulation and the resulting potential of mean force (PMF) would accordingly be dependent on the starting conformation of the system due to the so-called unrepresentative sampling.

To circumvent the problem, since the late 1960s, a number of sophisticated techniques and strategies have so far been proposed and advanced based on performing random walks using non-Boltzmann probability weight factors~\cite{16}. Replica-exchange umbrella sampling (REUS)~\cite{17} is a nearly recent of them, which combines replica-exchange (RE)~\cite{18} and umbrella sampling (US)~\cite{19} techniques, and therefore, has both merits of RE and US. The former belongs to the so-called generalized-ensemble algorithms~\cite{20}, while the latter is indeed a particular application of the more general importance sampling~\cite{21} used in statistics. In contrast to RE, all replicas in REUS have the same temperature value and this is the potential energy that is exchanged.

Based on the fact that US is a quasi-equilibrium method---in contrast to non-equilibrium approaches such as steered molecular dynamics (SMD)~\cite{22}, or targeted molecular dynamics (TMD)~\cite{23,24}---each replica in REUS has to be simulated and therefore equilibrated in the canonical ensemble independently. The problem arises when the reaction coordinate is defined as the distance between the two ends of the system, such as a single RNA or amyloid $\beta$-strand, and the associated PMF is to be calculated along that collective variable. In such a case, we may have several replicas, each of which has to be equilibrated independently in a way that the related end-to-end distance (or equivalently the two ends of the strand) must be kept fixed during the equilibration simulation. To overcome this problem, we carried out such a kind of constrained equilibration on a single, stretched $\beta$-strand of an amyloid beta dodecamer fibril using the constant-velocity protocol of SMD in that a differential value of the order of $10^{-20}$ \AA$\big/$fs was set to the pulling velocity of the SMD atom, which, in turn, kept this atom practically fixed over the total timespan. Examining the time-dependent total energies and forces acting on the SMD atom for all the replicas eventually verified the accuracy of the constrained equilibration. The computational setup was also described in Sec.~\ref{Sec:2} in detail.
\section{{\label{Sec:2}}Computational details}
The initial atomic positions of a single amyloid $\beta$-strand was taken from the RCSB~\cite{25} PDB file of amyloid beta A$\beta_{11-42}$ dodecamer fibril with the entry code 2MXU~\cite{26}. The simulations were carried out on Debian style~\cite{27} Linux~\cite{28} systems supported by MPICH (version 2-1.4)~\cite{29,30} using NAMD (version 2.14b2)~\cite{31} computer software with the July 2018 update of CHARMM36 force fields~\cite{32}. The VMD program (version 1.9.4a9)~\cite{33} was also used for post-processing. The fibril was solvated in a box of water under periodic boundary conditions with a unit-cell padding of about 2 nm to decouple the periodic interactions. Minimization was then carried out for 50000 conjugate gradient steps (100 ps) followed by a 2-ns free-dynamics equilibration at 310 K and 1.01325 bar in order to reach the equilibrium conformation of the fibril. Langevin forces with a Langevin damping constant of 2.5$\big/$ps along with the Langevin piston pressure control were applied for the NPT equilibration to keep fixed the temperature and pressure of the system. The integration timestep was also 1 fs. 

\begin{figure}[h]
	\centering
	\fbox{\rule[0cm]{0cm}{0cm} \rule[0cm]{0cm}{0cm}
		\subfigure[]{
		\includegraphics[scale=0.3]{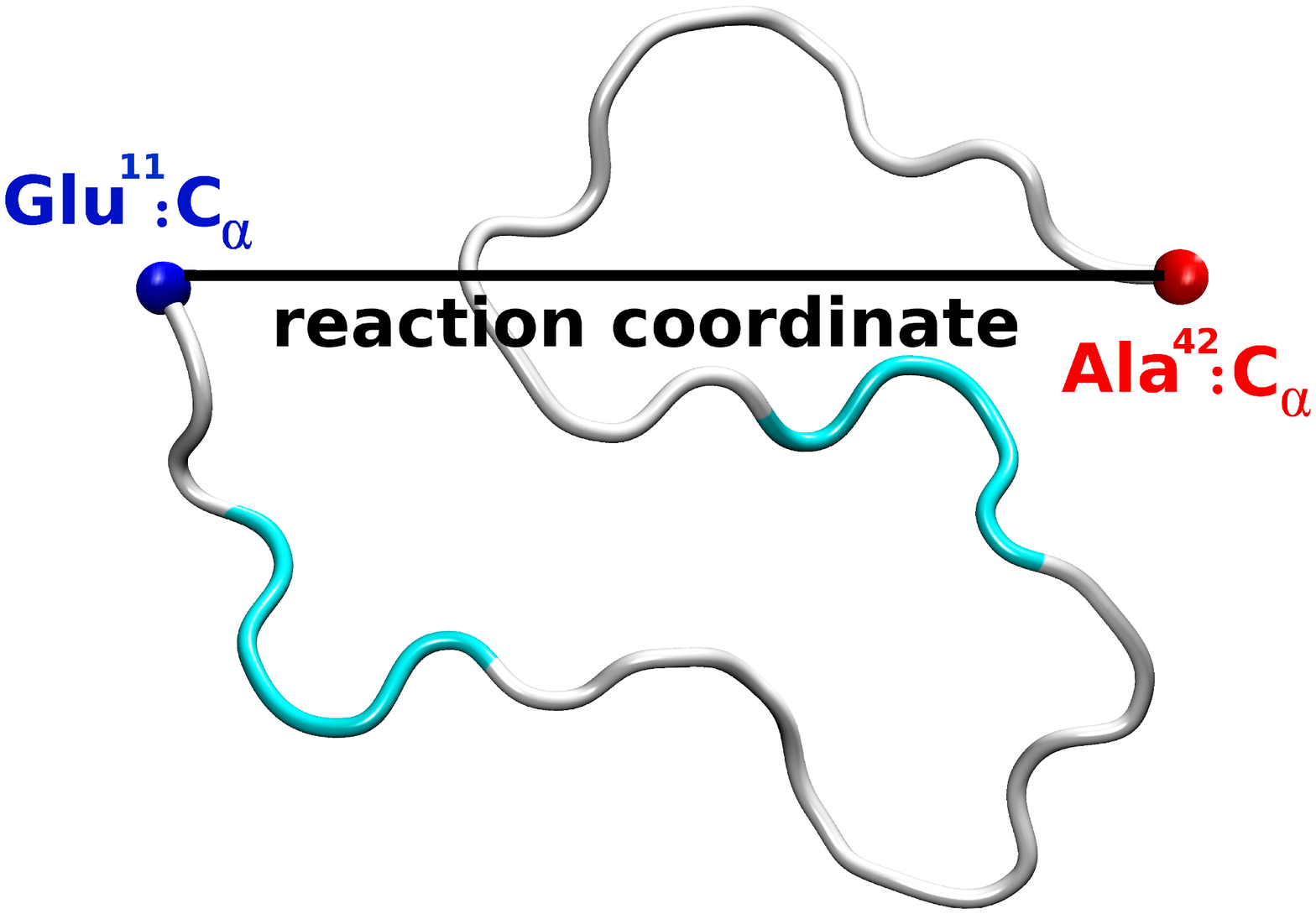}}
		\subfigure[]{
		\includegraphics[scale=0.42]{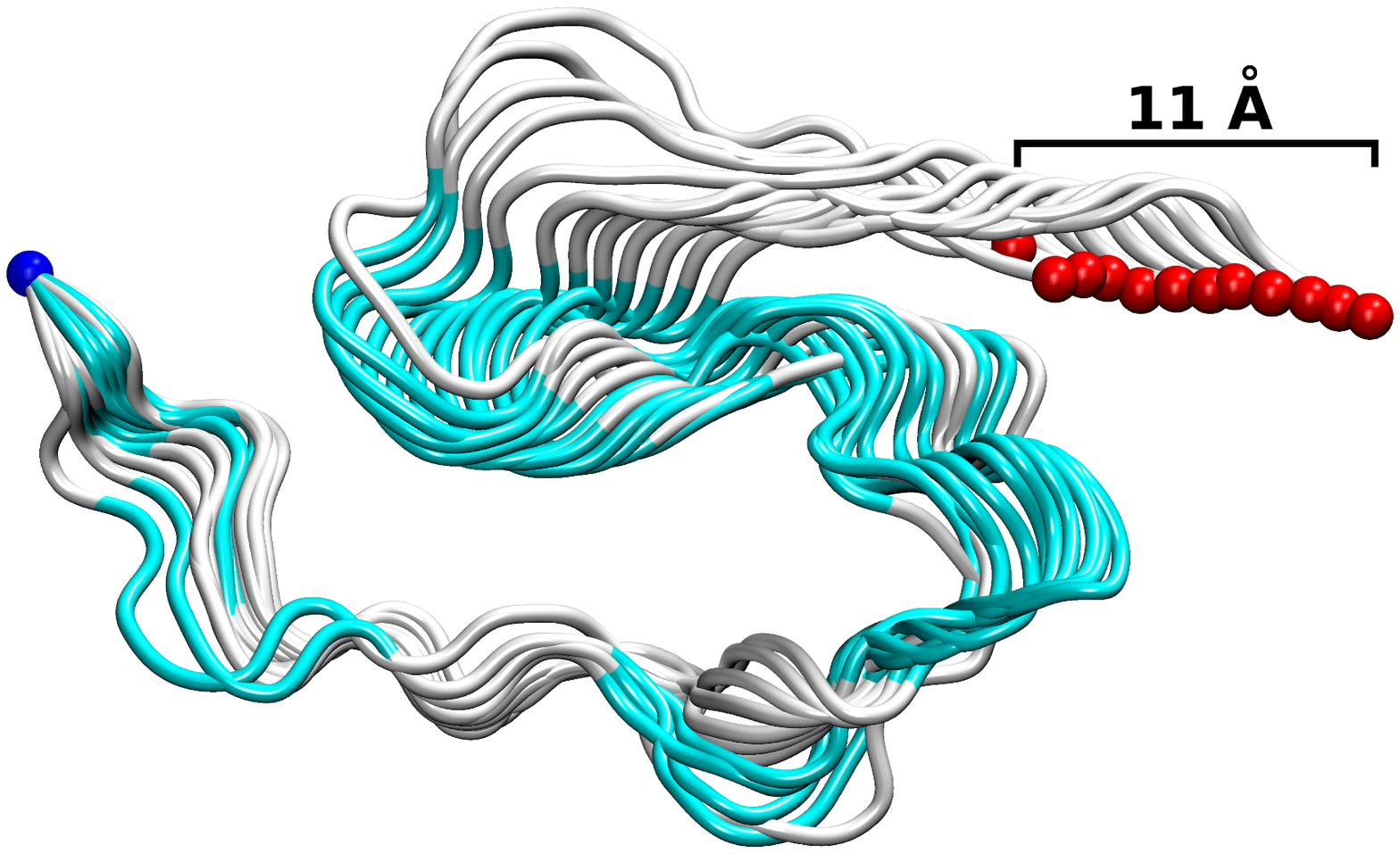}}}
	\caption{
		(a) The tertiary structure of a single $\beta$-strand taken from an S-shaped amyloid beta A$\beta_{11-42}$ dodecamer fibril and the associated reaction coordinate (collective variable) defined as the distance between Glu$^{11}$:C$_{\alpha}$ and Ala$^{42}$:C$_{\alpha}$---rendered in VMD using Tachyon parallel$\big/$multi-processor ray tracing system~\cite{34}. (b) Variations in the initial conformation of the $\beta$-strand during the 11-{{\AA}} traction in a way that each strand corresponds with the initial conformation of one replica. The fixed and SMD atoms at the two ends are shown by the blue and red balls, respectively.}
	\label{fig:1}
\end{figure}

After equilibrating the whole fibril in water, a single $\beta$-strand was taken and extended by SMD simulations in vacuum in a way that the distance between the first atom (Glu$^{11}$:C$_{\alpha}$; where C$_{\alpha}$ is carbon, 11 is the residue number, and Glu is the associated amino acid) and the last one (Ala$^{42}$:C$_{\alpha}$), defined as the reaction coordinate (Fig.~\ref{fig:1}), increased from 31.4 {{\AA}} (the initial value) to about 42.4 {{\AA}} according to the function $31.4+i{\hspace{1mm}}(i=0,...,11)$. Therefore, 12 independent replicas numbered consecutively from 0 to 11 were generated.
Each replica equilibrated for 1 ns at 310 K in vacuum during an SMD simulation in that Glu$^{11}$:C$_{\alpha}$ and Ala$^{42}$:C$_{\alpha}$ were respectively defined as the fixed and SMD atoms, and the latter was pulled along the reaction coordinate ($+z$; one-dimensional pulling) with an infinitesimal, constant velocity of $v_{p}=10^{-20}$ \AA$\big/$fs, which is practically zero over the total timespan. The force acting on the SMD atom was described as 
\begin{equation}
\label{eq:1}
{\bf{f}}({\bf{r}},t)=-{\bf{\nabla}} U({\bf{r}},t)=-{\bf{\nabla}}\left(\frac{k}{2}\big[v_{p}t-({\bf{r}}-{\bf{r}}_{0}).\hat{n}\big]^{2}\right), 
\end{equation}
where $U$ is the harmonic potential (spring), $k$ is the spring constant, $t$ is time, ${\bf{r}}$ and ${\bf{r}}_{0}$ are respectively the instantaneous and initial position vectors of the SMD atom, and $\hat{n}=(0,0,1)$ is the normalized pulling vector (along $+z$). The harmonic biasing potential of the form $U(\xi,i)=k(\xi-\xi_{i})^{2}\big/2$ was added to the Hamiltonian of the system, where $k=2$ kcal$\big/$(mol.{\AA}$^2$)---corresponding to a thermal fluctuation of the SMD atom of $\sqrt{k_{B}T/k}=0.555$ {\AA} at 310 K, with $k_B$ the Boltzmann constant---$\xi$ is the collective variable, and $\xi_{i}=(31.4+i)$ {\AA} defines the center of the $i$th bias potential.
\section{Results and discussion}
\begin{figure}[h]
	\centering
	\subfigure[replica $\#0$]{
	\includegraphics[scale=0.037]{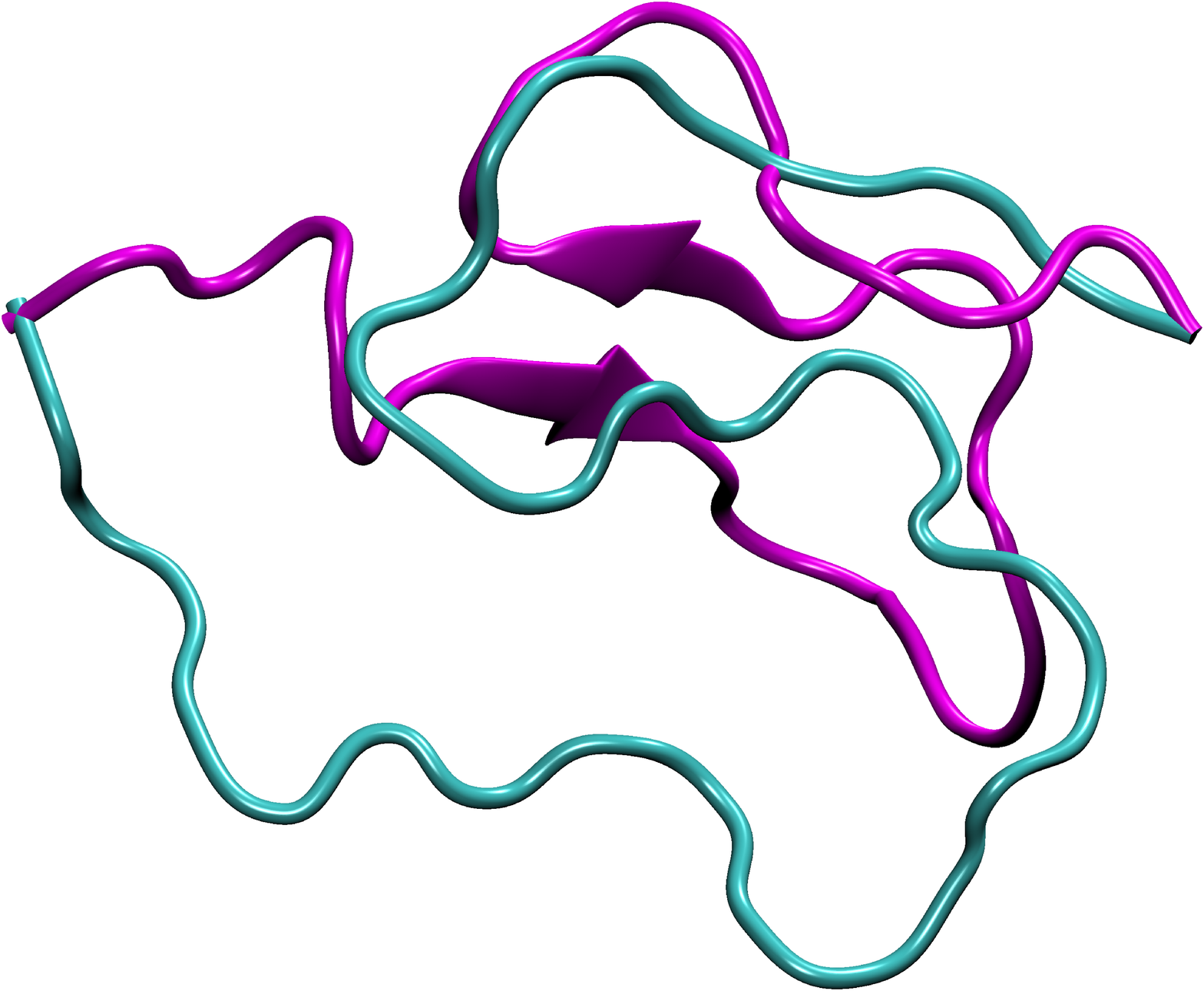}}
	\subfigure[replica $\#1$]{
	\includegraphics[scale=0.039]{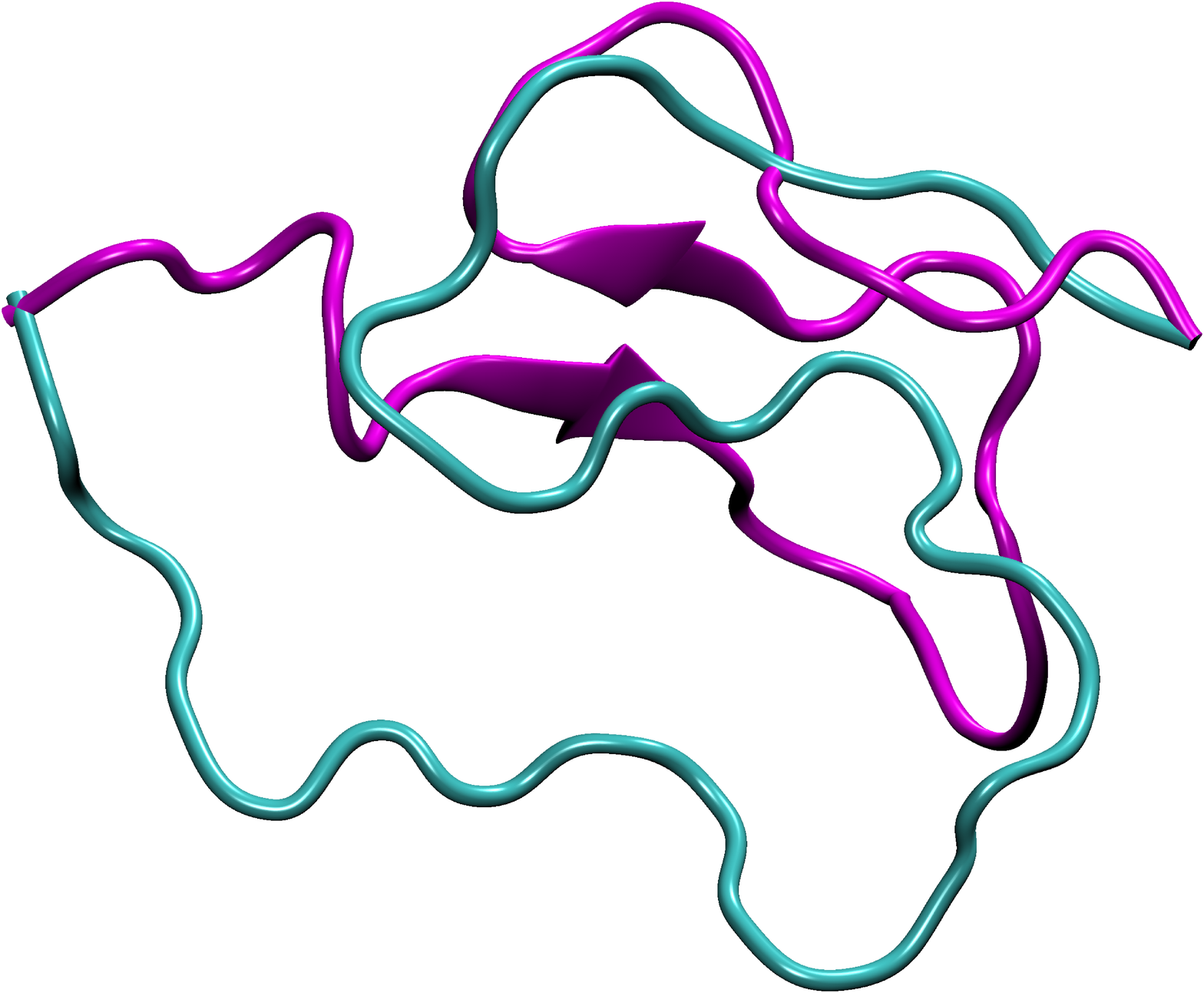}}
	\subfigure[replica $\#2$]{
	\includegraphics[scale=0.039]{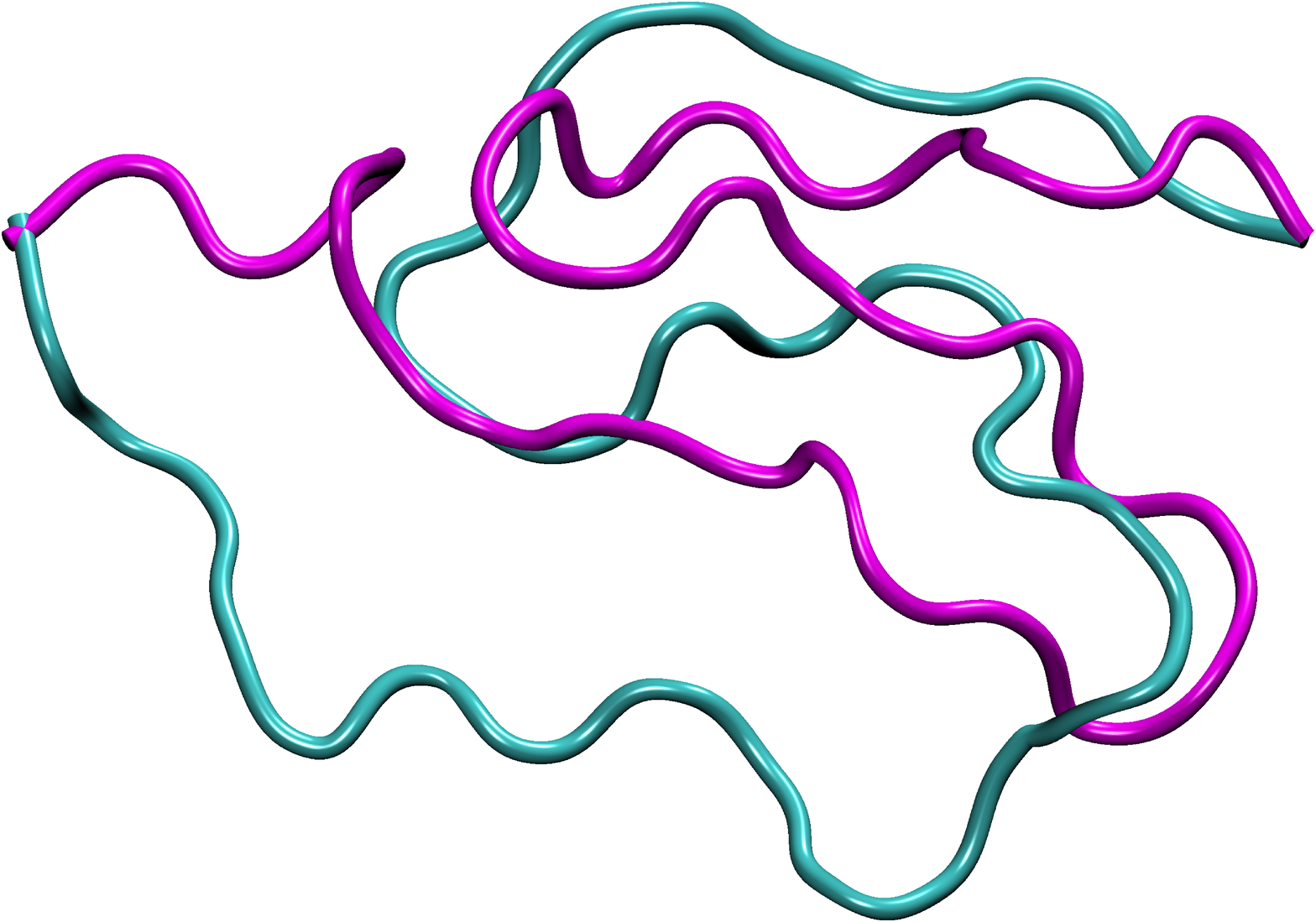}}
	\subfigure[replica $\#3$]{
	\includegraphics[scale=0.039]{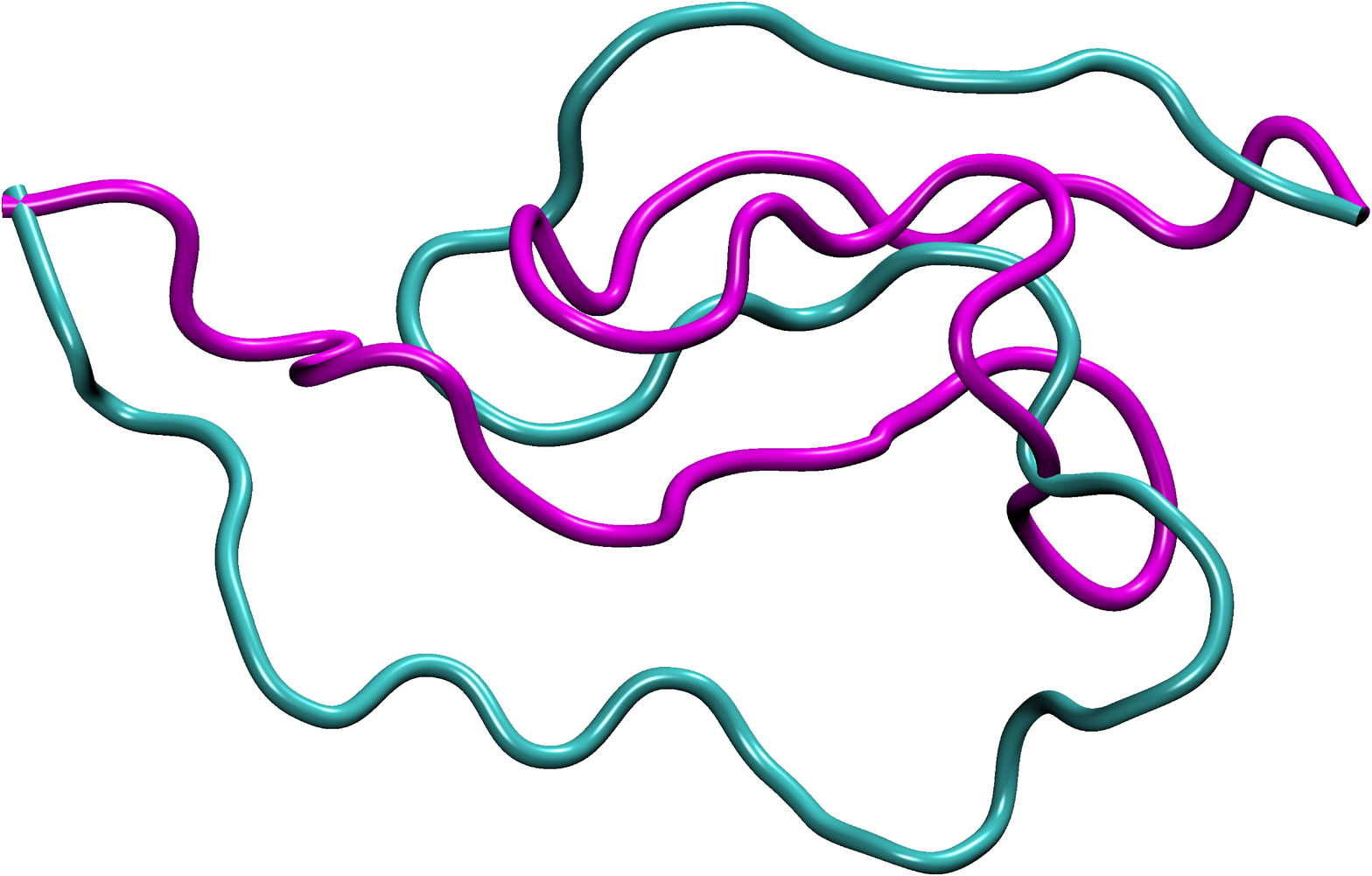}}
	\subfigure[replica $\#4$]{
	\includegraphics[scale=0.035]{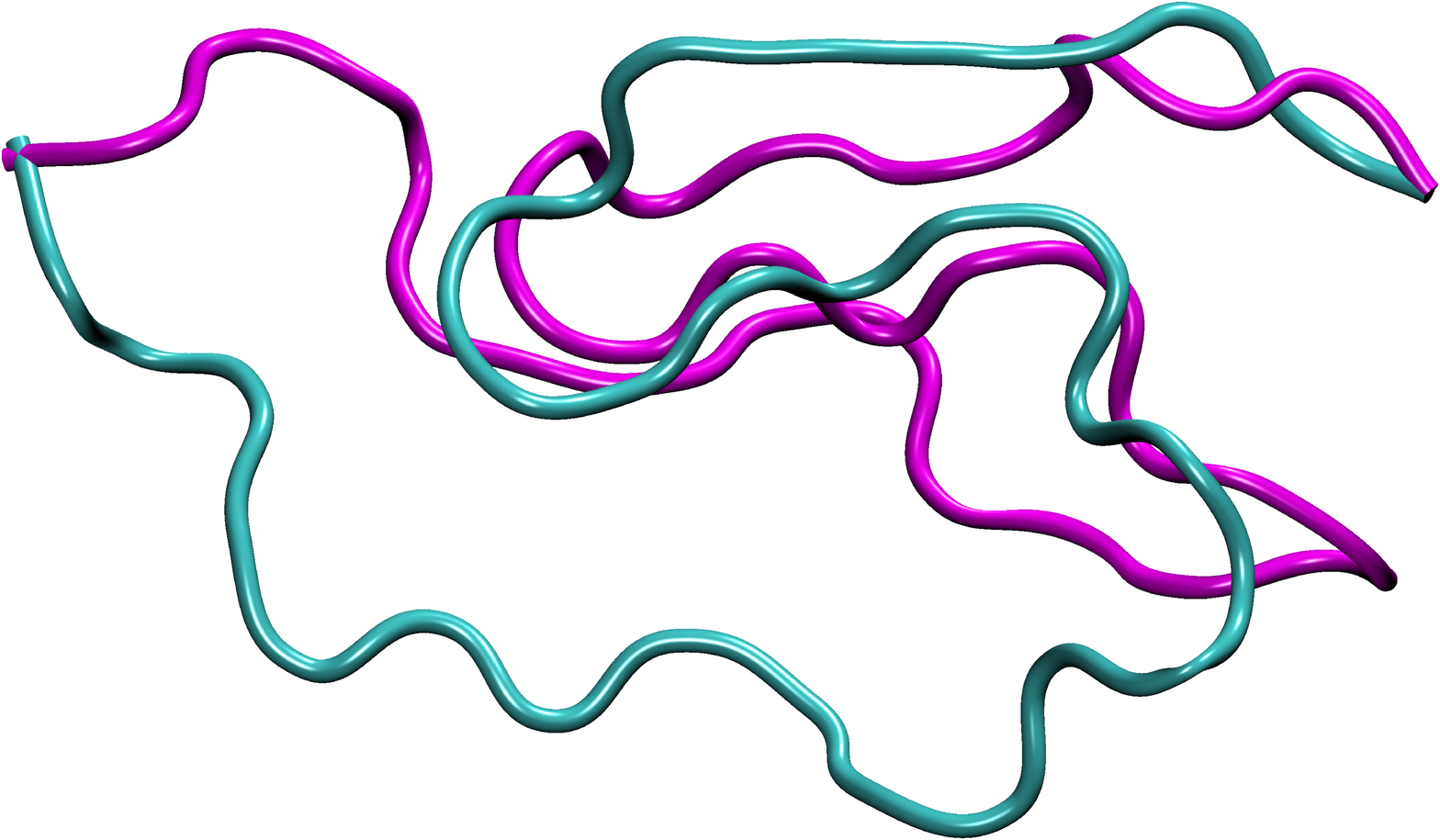}}
	\subfigure[replica $\#5$]{
	\includegraphics[scale=0.035]{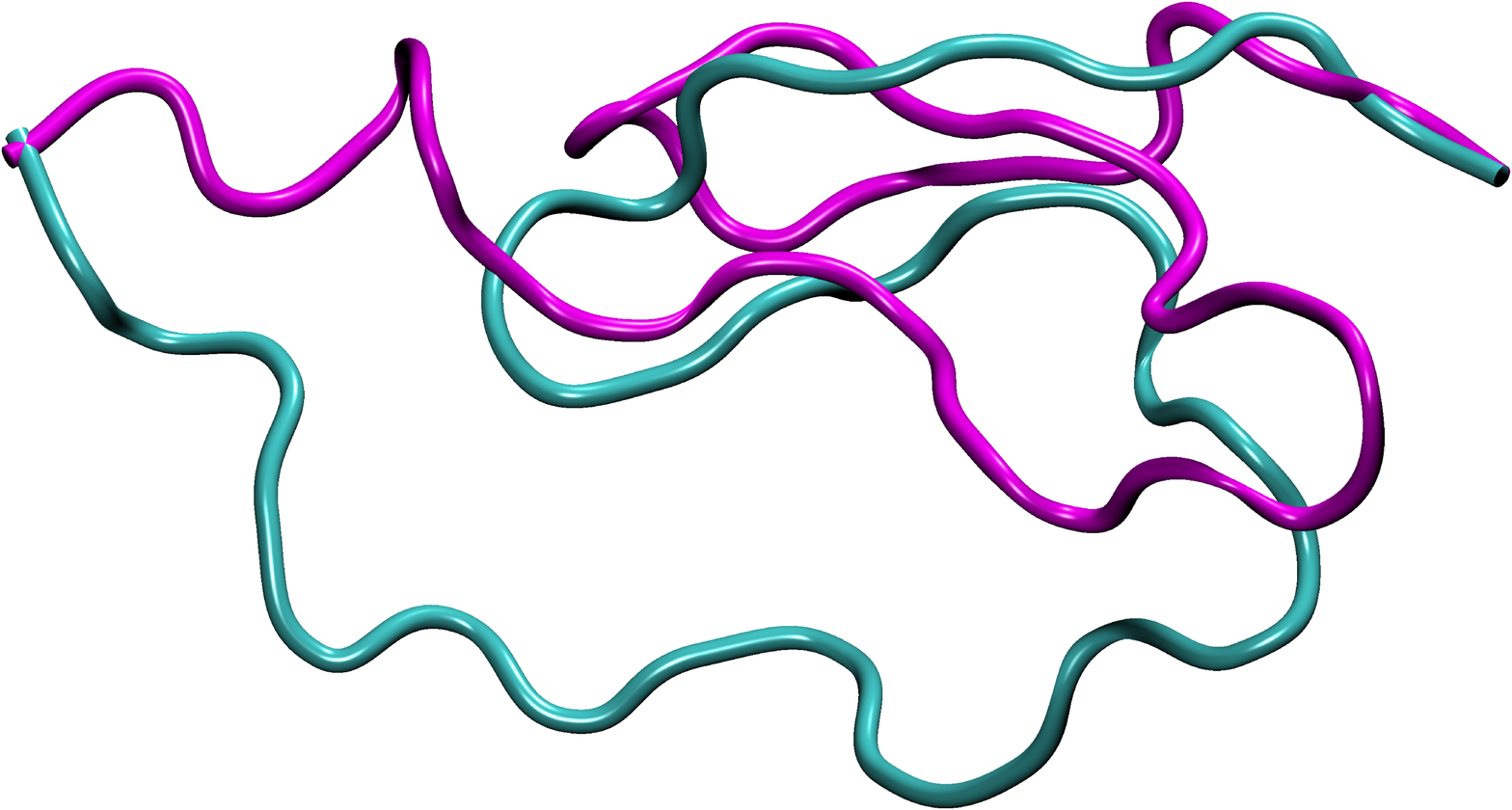}}
	\subfigure[replica $\#6$]{
	\includegraphics[scale=0.037]{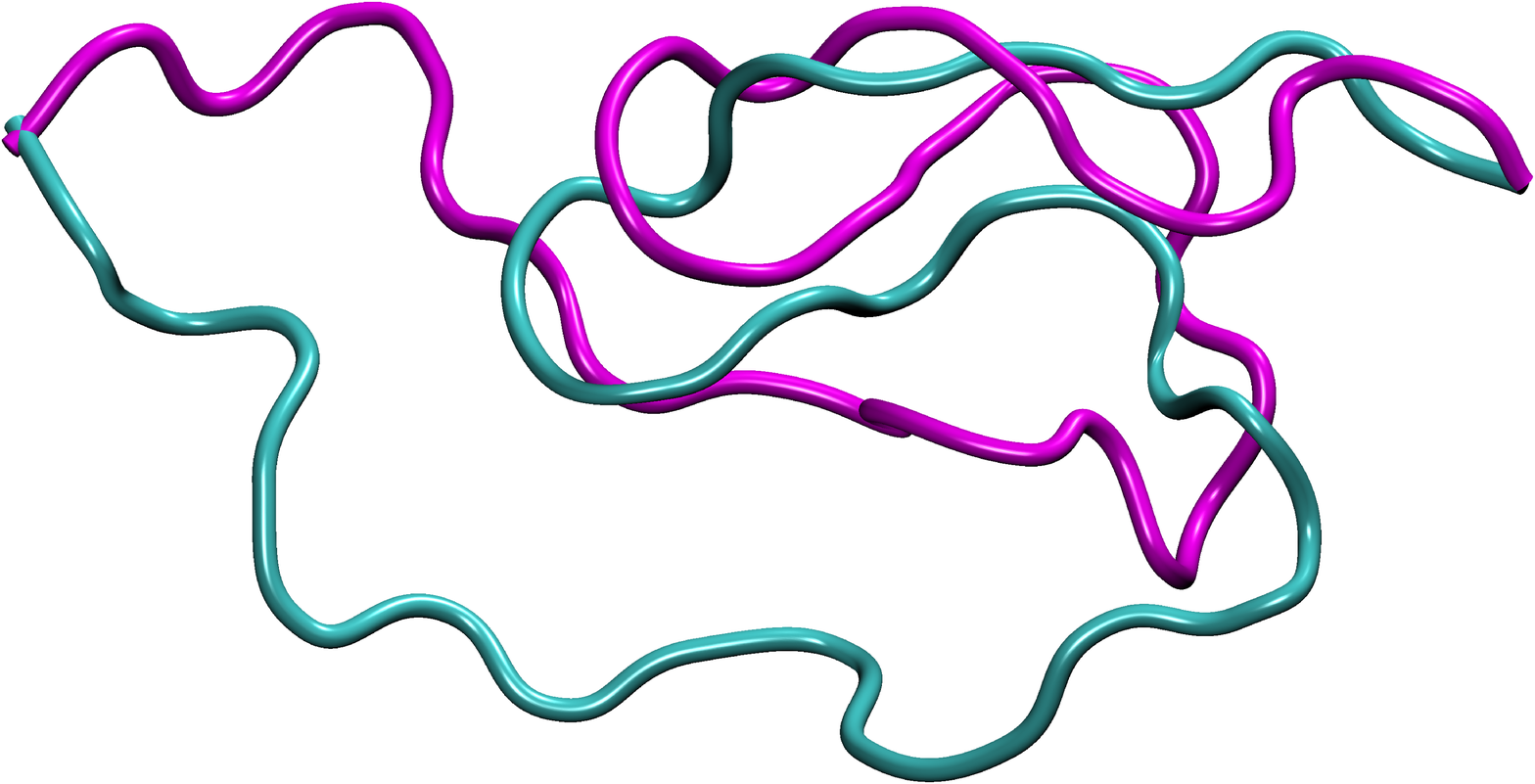}}
	\subfigure[replica $\#7$]{
	\includegraphics[scale=0.033]{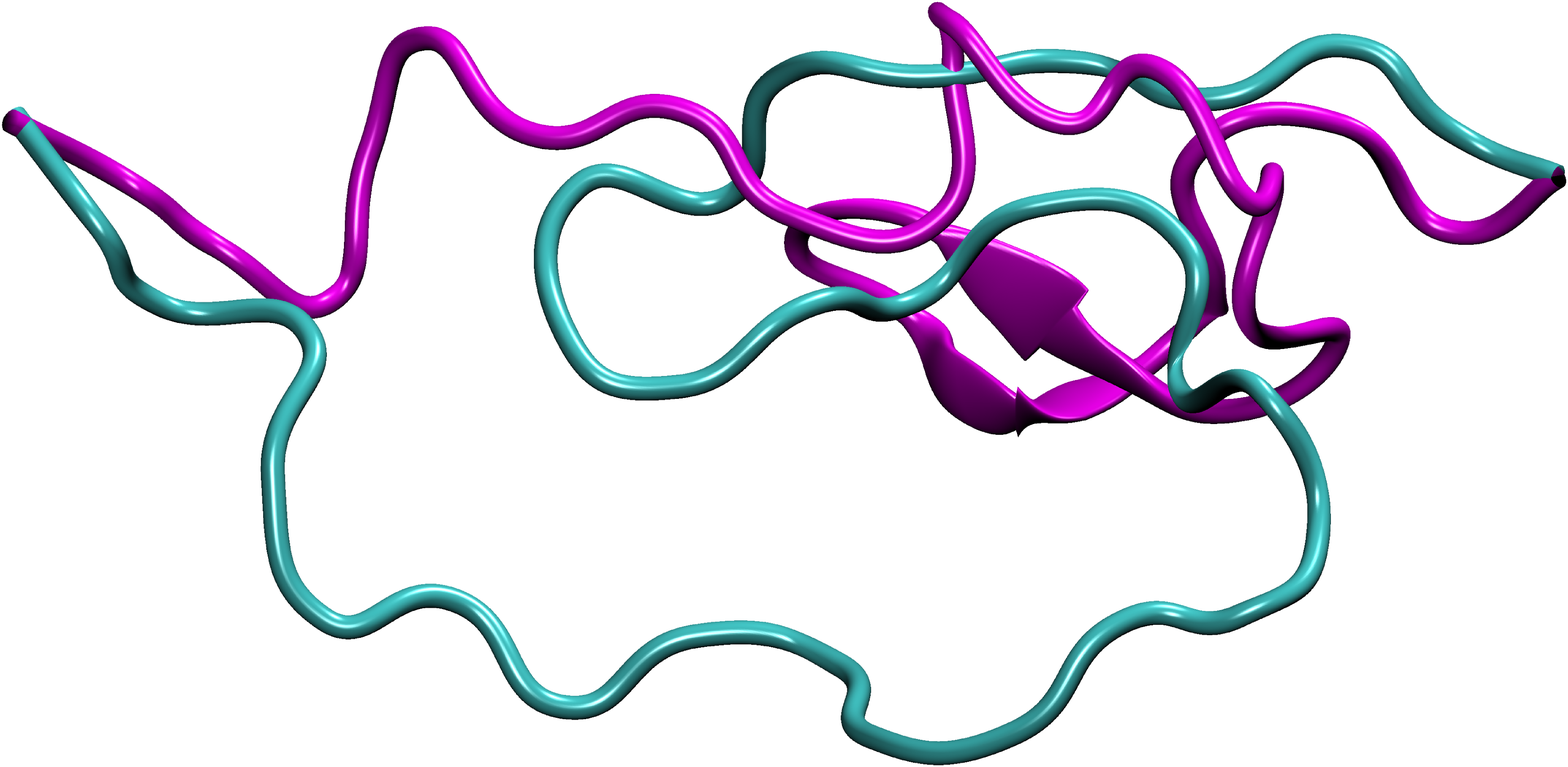}}
	\subfigure[replica $\#8$]{
	\includegraphics[scale=0.035]{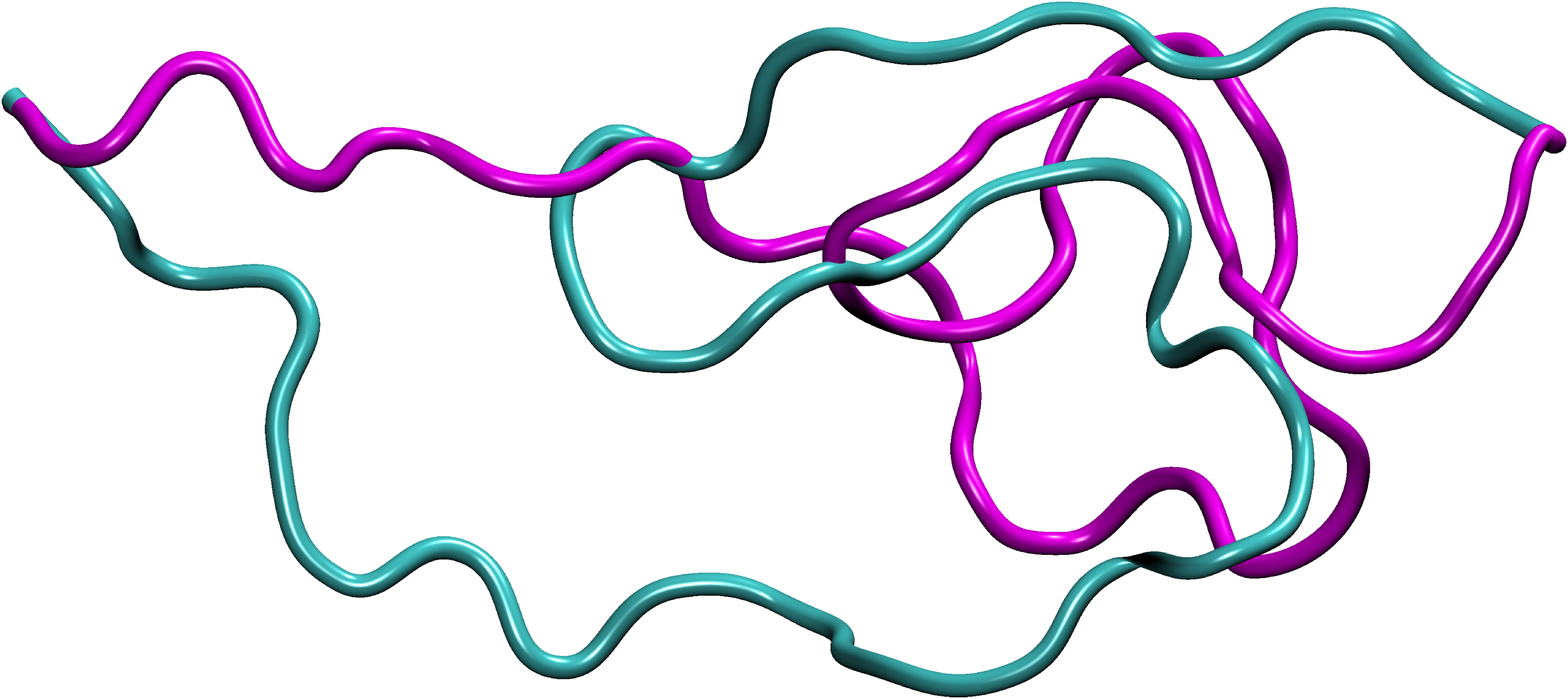}}
	\subfigure[replica $\#9$]{
	\includegraphics[scale=0.035]{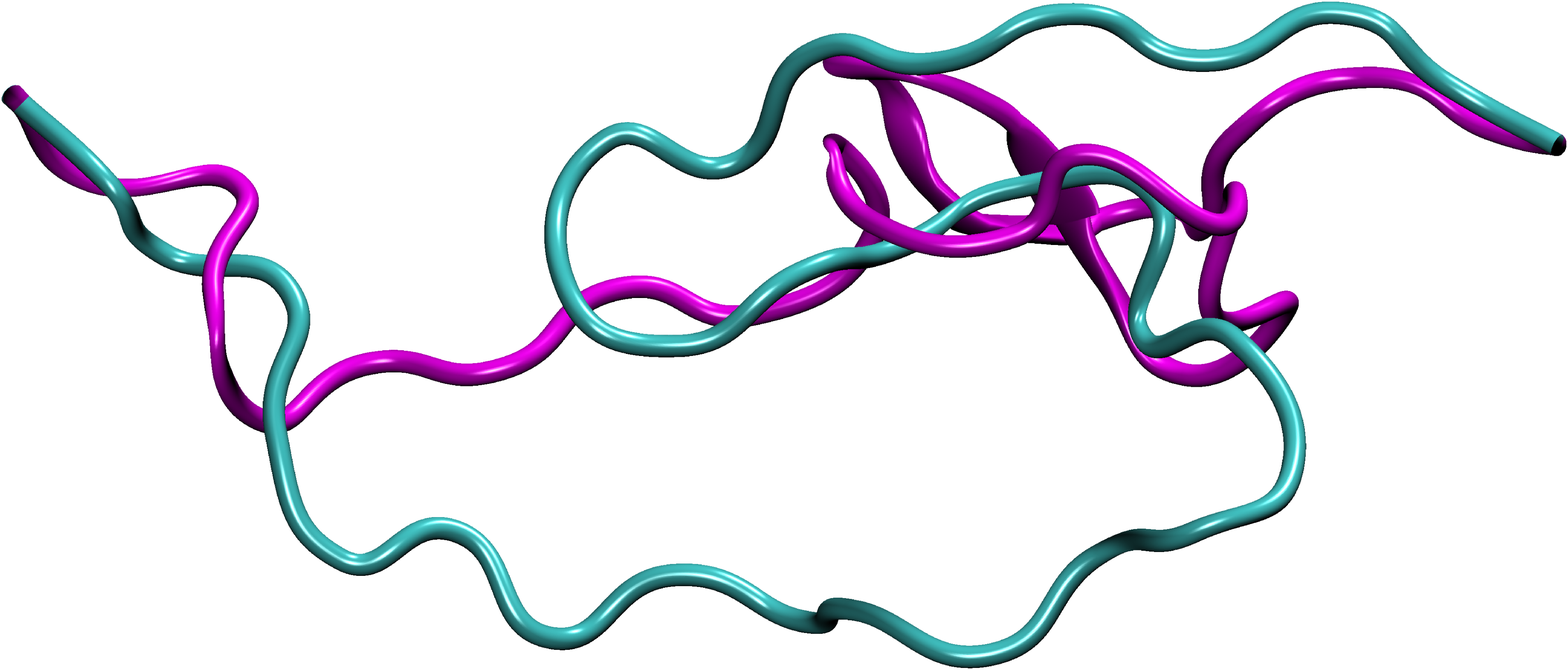}}
	\subfigure[replica $\#10$]{
	\includegraphics[scale=0.028]{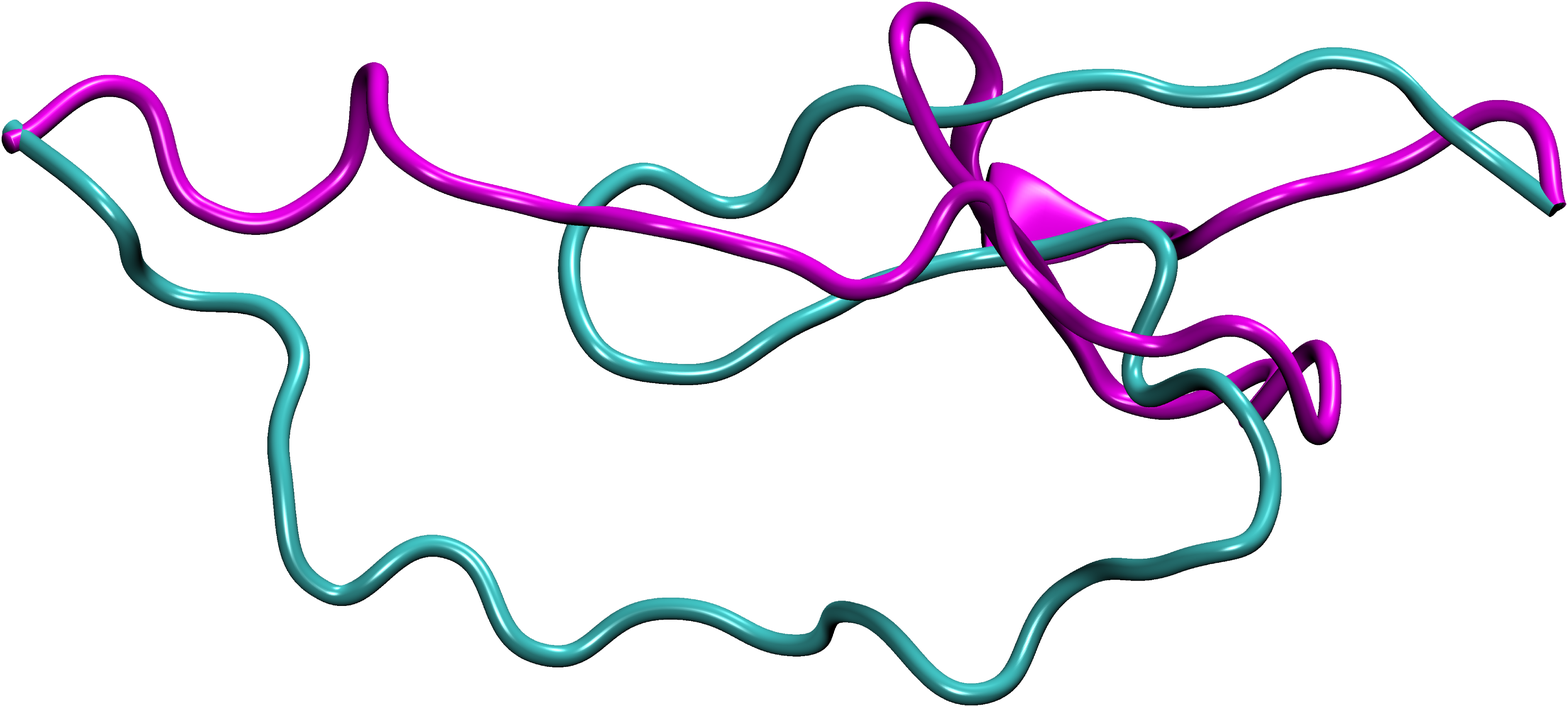}}
	\subfigure[replica $\#11$]{
	\includegraphics[scale=0.028]{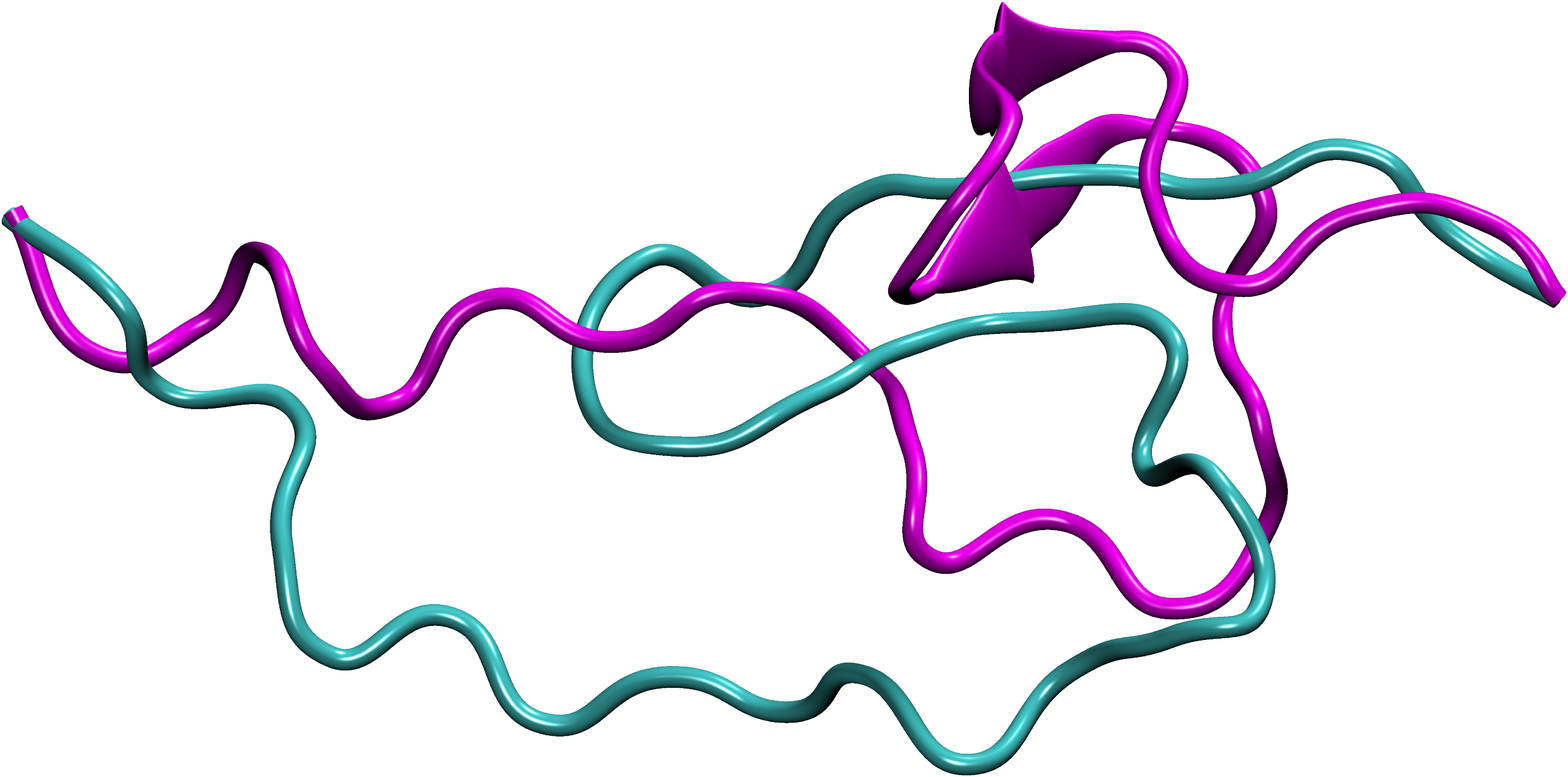}}
	\caption{
		Conformational transition of each replica from the associated initial state at $t=0$ (cyan) to the equilibrated conformation at $t=1$ ns (magenta). The right ends of the two strands for each replica exactly coincide, showing that the SMD atom is practically fixed during the simulations.}
	    \label{fig:2}
\end{figure}

\begin{figure}[h]
	\centering
	\fbox{\rule[0cm]{0cm}{0cm} \rule[0cm]{0cm}{0cm}
	\includegraphics[scale=0.632,angle=-90]{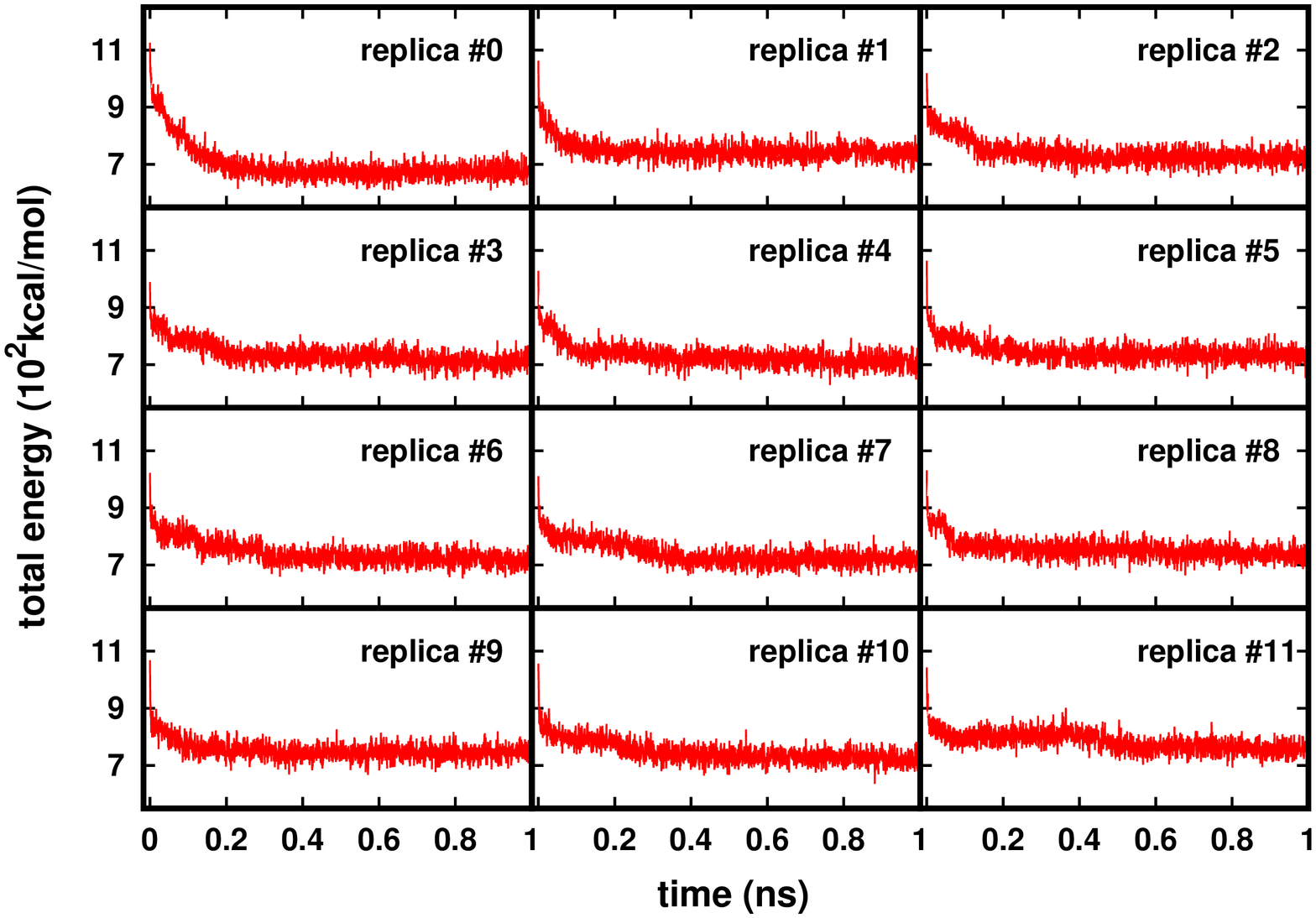}}
	\caption{
		The time-dependent total energy curves of the replicas during constrained equilibration---rendered in Gnuplot (version 5.2)~\cite{35}. The long-tailed converging patterns are evident, demonstrating that all the replicas are finally in equilibrium.}
	\label{fig:3}
\end{figure}

\begin{figure}[h]
	\centering
	\fbox{\rule[0cm]{0cm}{0cm} \rule[0cm]{0cm}{0cm}
	\subfigure[]{\label{subfig:4(a)}
	\includegraphics[scale=0.31,angle=-90]{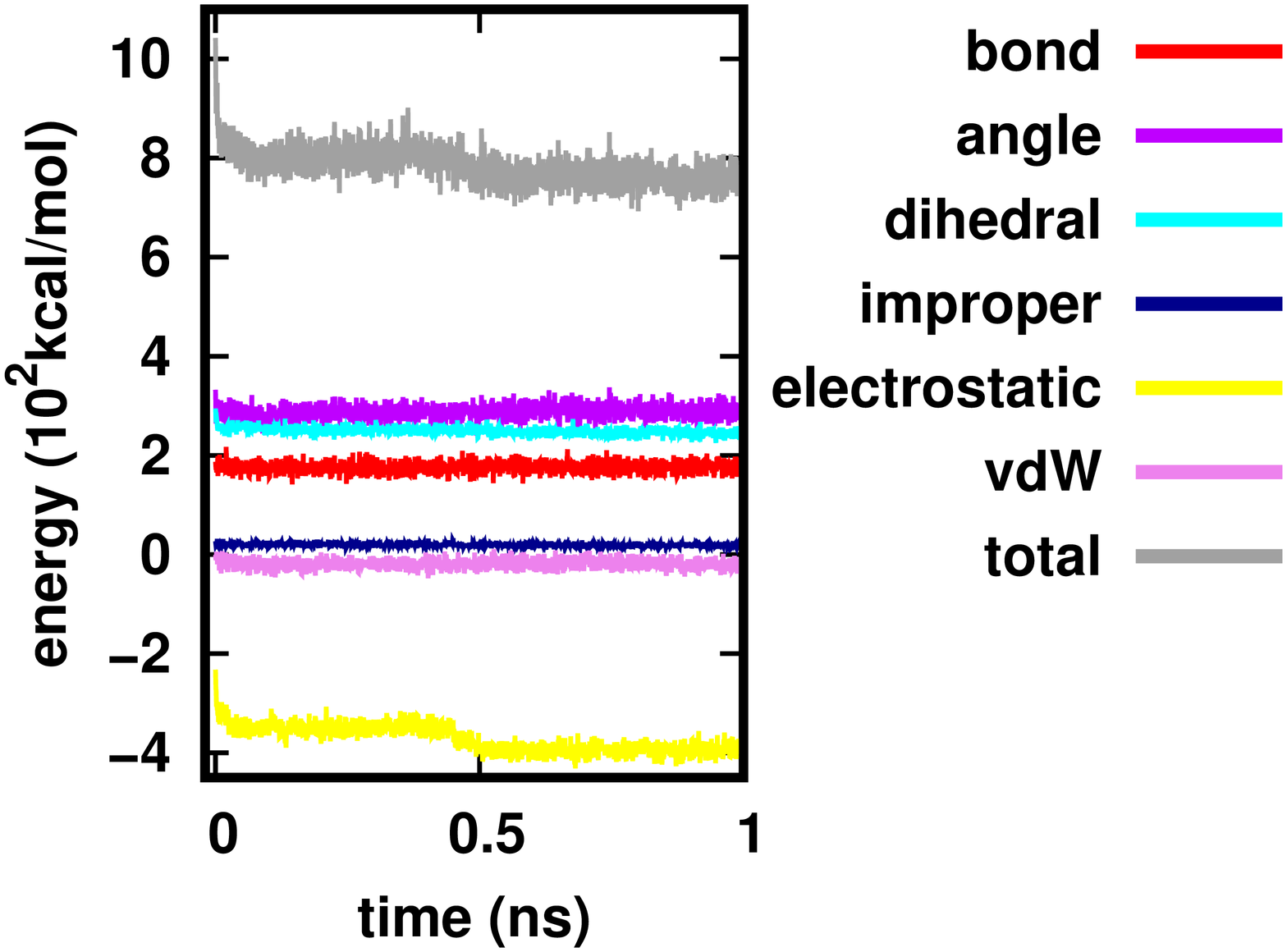}}
	\subfigure[]{\label{subfig:4(b)}
	\includegraphics[scale=0.31,angle=-90]{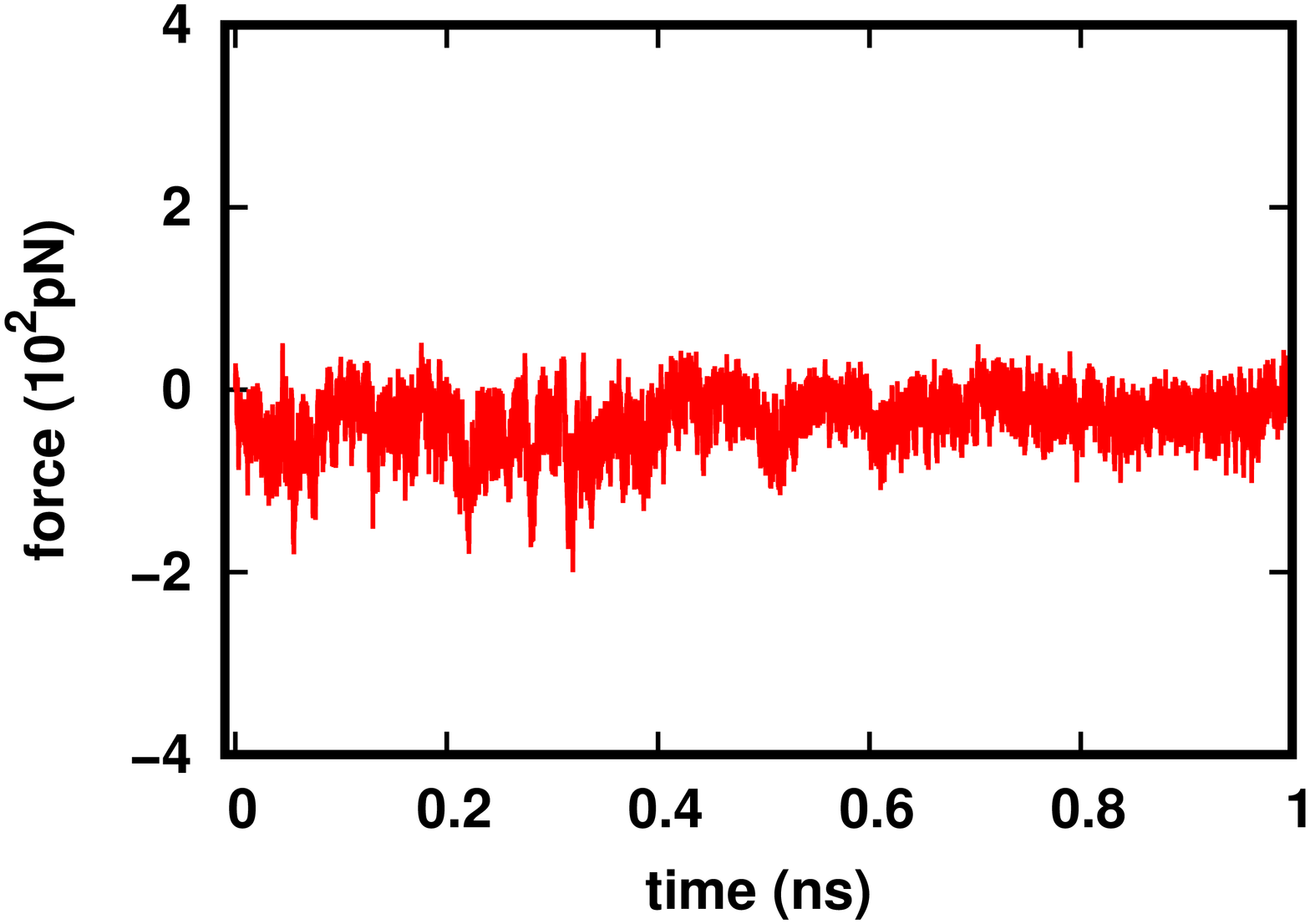}}}
	\caption{
		(a) The various energy contributions of the CHARMM potential energy function including bond, angle, dihedral, improper dihedral, electrostatic, van der Waals (vdW), and total energy, and (b) the associated instantaneous force exerted on the SMD atom, as functions of time over the total timespan of simulating the replica \#11. The average force is clearly zero over the last 0.5 ns due to achieving thermodynamic equilibrium.}
	\label{fig:4}
\end{figure}

Fig.~\ref{fig:2} illustrates the initial and equilibrated conformations of each replica shown in cyan (at $t=0$) and magenta (at $t=1$ ns), respectively.
As is seen, from replica \#0 to the last one, the end-to-end extension increases 1 {{\AA}} at a time, and each replica independently equilibrates at 310 K. Moreover, the right ends of the two strands for each replica exactly coincide, approving that the SMD atom, in practice, is fixed during the simulation, as expected. To verify whether the equilibrations have been achieved, the associated time-dependent total energy curves were accordingly provided, as shown in Fig.~\ref{fig:3}.
All the curves take similar, initially-decreasing trends and then converge to values around 700 kcal$\big/$mol with long tails, demonstrating that the corresponding equilibrium states have been attained. It could also be observed that the larger the end-to-end distance is stretched, the higher the value to which the total energy converges. Further examining the energy--time diagrams revealed that the specific pattern of the total energy was uniquely determined by that of the electrostatic energy based on the observation that the other energy terms in the CHARMM potential energy function---including bond, angle, dihedral, improper dihedral, and van der Waals (vdW)---fluctuated about constant values over the total timespan. This could be inferred from Fig.~\ref{subfig:4(a)}, which depicts the time dependence of these various energy contributions associated to the replica \#11.
Fig.~\ref{subfig:4(b)} also illustrates the time dependence of the instantaneous force exerted on the SMD atom during the constrained equilibration of the replica \#11. The relatively-large fluctuations over the first half of the trajectory are mainly due to the fact that the strand is initially out of equilibrium, and therefore considerable forces act on the SMD atom over this period, in contrast to the second half over which equilibration is achieved and the fluctuating pattern then becomes narrower. That these fluctuations mostly take negative values over the first 0.5 ns further reveals that the force is exerted by the strand itself, not by the external pulling factor which is practically vanishing during the simulation according to Eq.~\ref{eq:1} and $v_{p}=10^{-20}$ {{\AA}}$\big/$fs. The average value of the force over the second half of the trajectory is also clearly zero due to equilibration as well as the infinitesimal pulling velocity.
\section{Conclusions}
We proposed a simulation trick to carry out equilibration simulations on biological macromolecules, which are constantly under the action of externally-applied, mechanical constraints such as stretching or contraction. Such cases could frequently be encountered in replica-exchange umbrella sampling simulations particularly where the reaction coordinate is explicitly related to the mechanical deformation of the system and the potential of mean force (PMF) is the dependent quantity of interest. We accordingly applied our approach to equilibrate a single, stretched $\beta$-strand, taken from an amyloid beta dodecamer fibril, where the reaction coordinate was defined as the distance between the two ends of the strand, and was accordingly extended by 11 {{\AA}} in a way that every 1 {{\AA}} of extension generated the initial conformation of one replica, leaving 12 replicas as well. We then simulated each replica independently in the canonical ensemble at 310 K using constant-velocity SMD in a way that the SMD atom was being pulled with a differential velocity of magnitude $10^{-20}$ \AA$\big/$fs along the reaction coordinate, making this atom to be practically fixed during the simulations. As a result, the SMD simulation turned into an equilibrating process and each replica consequently equilibrated using free dynamics plus two mechanical constraints corresponding with the two fixed ends. The validity of this kind of constrained equilibration was strongly approved by examining the time-dependent variations of the total energy and the force exerted on the SMD atom associated to each replica.
\bibliographystyle{unsrt}

\end{document}